\documentclass[10pt, conference, letterpaper]{IEEEtran}
\IEEEoverridecommandlockouts
\usepackage{amsmath,amssymb,amsfonts}
\usepackage{graphicx}
\usepackage{textcomp}
\usepackage{xcolor}
\usepackage{pifont}

\usepackage{flushend}
\usepackage{paralist}
\usepackage[abbreviations]{foreign}  

\usepackage{algorithm}
\usepackage{algorithmicx}
\usepackage{comment}
\usepackage{algpseudocode}
\usepackage{subcaption}

\newcommand{\lz}{$LØ$}

\usepackage{amssymb} 
\usepackage{array}   
\usepackage{tabularx} 

\usepackage{xspace}
\newcommand{\protocol}{\textsc{Hermes}\xspace}

\graphicspath{{figures/}}

\usepackage[colorlinks,
citecolor=violet,
urlcolor=black,
linkcolor=violet,
bookmarks=false,
hypertexnames=true]{hyperref}

\usepackage[
  sortcites,
  sorting=none,
  style=ieee,
  backend=biber,
  isbn=false,
  url=false,
  doi=false,
  url=false,
  eprint=false,
  mincrossrefs=100,
  maxnames=12,
]{biblatex}

\AtEveryBibitem{\clearname{editor}}

\begin{document}



\title{Mitigating Front-Running Attacks through Fair and Resilient Transaction Dissemination
    \thanks{This research was in part funded by the French National Research Agency (ANR) under project number ANR-21-CE25-0021-03 (GenBlock) and by the Luxembourg National Research Fund (FNR), through grant reference C22/IS/17232062 (SpaceVote). In fulfillment of the obligations arising from the grant agreement, the author has applied a Creative Commons Attribution 4.0 International (CC BY 4.0) license to any Author Accepted Manuscript version arising from this submission.}}

\author{
    \IEEEauthorblockN{Wassim Yahyaoui$^1$, Joachim Bruneau-Queyreix$^2$, Jérémie Decouchant$^3$, Marcus Völp$^1$}
    \IEEEauthorblockA{
        $^1$ Univ. of Luxembourg, SnT - CritiX group \qquad $^3$ Delft University of Technology\\ $^2$ Univ. Bordeaux, CNRS, Bordeaux-INP, LaBRI UMR5800, F-33400 Talence, France\\
        \url{wassim.yahyaoui@uni.lu},
        \url{joachim.bruneau-queyreix@u-bordeaux.fr},
        \url{j.decouchant@tudelft.nl},
        \url{marcus.voelp@uni.lu}
    }
}

\maketitle

\begin{abstract}
    In modern blockchains, efficient, fair, and fault-tolerant information dissemination is critical for performance and security. Several stages of the transaction lifecycle are affected, from the creation and dissemination of transactions to the dissemination of blocks in the consensus layer. Mempool protocols, such as \lz, already address some of modern blockchains’ security threats. However, others remain unless fairness is embedded most fundamentally in the dissemination layer used by these protocols to share transactions and blocks and to reconcile mempools.
    This paper introduces \protocol{}, a novel dissemination protocol for mempools that leverages robust minimal structures to optimize data propagation, balance load, and ensure fairness despite faults and Byzantine actors.
    Specifically, we address front-running attacks by randomizing the choice of an overlay structure while forcing nodes to prove adherence to the mempools' dissemination policies and the random choice made.
    Experimental results show significant performance improvements compared to traditional broadcast protocols for both permissioned and permissionless blockchains.
\end{abstract}


\section{Introduction} 
\label{sec:introduction}

Efficient, fair, and accountable information dissemination despite accidental and intentionally malicious node and link failures remains a challenging problem. This is particularly true for both permissioned and permissionless blockchains, due to their increasing complexity and scale.
Node failures, leading to inefficient bandwidth utilization and repeated retransmission, threaten these properties, and unfair or unbalanced dissemination give rise to security threats, such as miner exploitable value (MEV) attacks.
For example, in 2013, Decker et al.~\cite{decker2013information} proved that propagation delays are the primary cause of blockchain forks, which makes the blockchain vulnerable to double spending and block-withholding attacks. In 2020, Mao et al.~\cite{mao2020perigee} highlighted the importance of block propagation delays for the performance of blockchain networks.
Currently employed dissemination strategies, including gossip, give rise to front-running attacks leading to unfair competitive advantages
when disseminating transactions. The root cause for these attacks is that some nodes may learn about transactions earlier than others.
Nodes can exploit this knowledge if they can control how their transactions are disseminated in relation to when they receive this information.
Furthermore, nodes can be systematically overwhelmed by a flood of dissemination requests.
In such situations, excessive reliance on retransmission further strains network resources, leading to inefficiencies that degrade bandwidth and increase latency. In some situations, flooding can also open the door to further attacks in the worst scenarios.

\lz{}~\cite{nasrulin2023accountable} already addresses a large class of MEV attacks by introducing accountability in the dissemination layer. In a nutshell, \lz{}'s miners generate, exchange with each other, and record cryptographic commitments of the transactions they know (i.e., their mempools) before receiving transactions from each other. Miners then witness each other's behavior, including how their peers disseminate transactions and bundle them into the blocks they create. Because commitments are further exchanged among miners, \lz{} uncovers reordering attacks with high probability. Unfortunately, since the dissemination paths in \lz{} remain to a large degree in the hands of miners (they can establish additional links on top of the singular unidirectional overlay), they may exploit early knowledge by choosing dissemination paths that favor the miner over other nodes. The result is a front-running attack.

Of course, preventing nodes from learning information early is difficult, if not impossible.
For example, malicious nodes may well tap into the protocol implementation and access information when received, rather than when delivered.
Moreover, they may establish channels outside the overlay, leveraging links in the physical network, or even establish additional networks.
We therefore have to tackle the problem by preventing miners from exploiting this advantage. We do so by preventing them from controlling the order in which related messages are delivered.

We present \protocol, a protocol that extends \lz{} to prevent font-running and ensures \emph{dissemination-fair}, accountable, Byzantine-resilient, and efficient transaction dissemination for blockchain systems.
To be practical, \protocol also maintains a low network overhead and a low computational overhead.
By \emph{dissemination fairness}, we mean ensuring that no correct node is systematically left behind in the delivery of messages and no adversary can exploit timing advantages to reorder messages or front-run their dissemination.
Whether by overloading specific nodes to delay message propagation or by acquiring early knowledge on the message dissemination, our goal is to prevent adversaries' front-running attempts.

To achieve this, \protocol constructs multiple overlay structures optimized for robust and efficient propagation.
\protocol{} simply modifies the neighbor selection procedure of \lz{} by selecting the children of a node in its precomputed dissemination overlays.
Having analyzed four dissemination structures, we found robust-tree topologies to yield minimal latency.
We have therefore selected robust trees as our overlay topology, but apply simulated annealing to systematically prune links and reposition nodes to optimize for low latency, and balance the dissemination roles, while preserving $f{+}1$ connectivity.
Nodes participate in all overlays.
By balancing dissemination roles we ensure that nodes near the root in one overlay, which naturally receive messages sent through that overlay before further away nodes, assume a more distant role in other overlay structures.
To prevent malicious nodes from predicting message dissemination times and gain a persistent advantage over others, which would allow them to successfully front-run transactions, \protocol enforces a randomized selection of the overlay a sender must use, leveraging a novel Threshold Random Seed (TRS) scheme.


We further enforce accountability by requiring nodes to prove that they have relayed messages correctly in the designated overlay.
Combined with thorough logging to trace node activity, \protocol prevents front-running attempts from remaining undetected, which allows identifying and excluding deviant nodes from the system.

We combine, for the first time, accountability at the mempool layer with randomized dissemination over optimized overlay structures.
More specifically, \protocol makes the following contributions:

\begin{itemize}
    \item \textbf{Low-latency balanced overlays:} We present a method for constructing a set of $k$ \emph{robust-tree} overlays and optimize them via simulated annealing. The latter  prunes redundant links and balances node roles across overlays to reduce dissemination latency, avoid bottlenecks at frequently used nodes, and provide fair dissemination.
          Further, each overlay structure is encoded, signed by a $3f{+}1$ committee, and disseminated to all nodes.
    \item \textbf{Threshold random seed (TRS) construction:}
          At the core of \protocol’s front-running resistance is our TRS construction.
          Our approach leverages reliable broadcast among a committee of $3f{+}1$ nodes to generates  a reliable seed for each dissemination round.
    \item \textbf{Randomized and accountable overlay selection:} Via our TRS construction, each message is verifiably bound to a randomly chosen overlay, preventing senders from picking routes that favor them.
          This ensures dissemination-fairness, enforces adherence to the assigned path, and eliminates systematic front-running attacks.
    \item \textbf{Robust and accountable dissemination:}
          Our design ensures resilience with $f{+}1$ connectivity per overlay, tolerating up to $f$   faulty nodes locally to prevent isolation or censorship.
          Dissemination accountability is enforced by relying on the signed overlay structures and via message sequencing: nodes verify senders as legitimate predecessors in their overlay, and are required to propagate missing messages before relaying their $i^{th}$ message, preventing selective omission.
\end{itemize}


Section~\ref{sec:overview} explains \protocol in a nutshell. Section~\ref{sec:overlay} describes our method for overlay construction, while Section~\ref{sec:account-fair} explains how \protocol achieves dissemination-fairness and accountability.
We analyze the performance of our protocol in Section~\ref{sec:evaluation}.
We start by establishing the above in permissioned settings, where network topologies are known. In Section~\ref{sec:dynamic_topologies}, we relax these constraints to permissionless settings with changing but generally sufficiently stable topologies.

\section{Related Work} 
\label{sec:related_work}

\textbf{Propagation Delay.}
Decker et al.~\cite{decker2013information} and Mao et al.~\cite{mao2020perigee} highlight the importance of propagation delay for the efficiency and security of blockchains.
Perigee~\cite{mao2020perigee} dynamically adjusts node connections to optimize propagation times.
CougaR~\cite{kolyvas2022cougar} employs random link selection to balance network latency and protect against eclipse attacks.
BlockP2P~\cite{hao2019blockp2p} accelerates broadcast by geographically clustering.
BlockP2P-EP~\cite{hao2020towards} builds on this by using overlay networks to further enhance performance.
Mercury~\cite{zhou2023mercury} leverages virtual coordinates and an early outburst strategy to reduce transaction latency.
Our protocol creates a set of latency-optimized robust overlays and randomly selects for each sender and message the ones to use in order to mitigate front-running attacks.



\textbf{Overlay Topologies.}
Several works leverage overlay structures in Blockchains.
For example, Kauri~\cite{neiheiser2021kauri}, Byzcoin~\cite {kogias2016enhancing}, Motor~\cite{kokorisrobust}, and Omniledger~\cite{kokoris2018omniledger} leverage tree structures to minimize latency and balance load.
SplitStream~\cite{DBLP:conf/iptps/CastroDKNRS03} addresses link failures by distributing load across multiple trees.
Ramanathan et al.~\cite{ramanathan1988reliable} and Toshniwal et al.~\cite{toshniwal2021comparative} use hypercubes and show that they outperform fat-tree, torus and scale-free technologies. You et al~\cite{youhyperclique} overcomes the power of two limitations of hypercubes by constructing hyperclique overlays.
Gossip protocols~\cite{DBLP:conf/ccs/ZamaniM018,DBLP:conf/aft/RohrerT19,DBLP:conf/p2p/FreyGKM10} avoid some of the flooding-related overheads. Bitcoin and, at least partially, Ethereum~\cite{ethereum2022} leverage gossip to redistribute learned information up to a certain age. Gossip achieves scalability and robustness to node and link failures, albeit at the costs of guaranteeing delivery only with a certain probability.
Our protocol constructs robust trees as overlay structures.

\begin{table*}[t]
    \caption{Comparison of transaction dissemination approaches}
    \label{tab:metrics_comparison}
    \centering
    \begin{tabular}{|>{\centering\arraybackslash}p{0.11\textwidth}|>{\centering\arraybackslash}p{0.18\textwidth}|>{\centering\arraybackslash}p{0.18\textwidth}|>{\centering\arraybackslash}p{0.18\textwidth}|>{\centering\arraybackslash}p{0.18\textwidth}|}
        \hline
        \textbf{Metric}         & \textbf{Gossip}      & \textbf{Reliable Broadcast} & \textbf{Simple Tree} & \textbf{\protocol{} (Robust Trees)} \\
        \hline
        Dissemination Mechanism & Direct communication & Randomized gossip           & Fixed tree overlays  & Optimized robust tree overlays      \\
        \hline
        Latency                 & Moderate             & Low                         & High                 & Moderate                            \\
        \hline
        Msg Complexity          & Moderate             & High                        & Low                  & Moderate                            \\
        \hline
        Dissemination-fairness  & \checkmark           & \ding{54}                   & \ding{54}            & \checkmark                          \\
        \hline
        Accountability          & \ding{54}            & \checkmark                  & \ding{54}            & \checkmark                          \\
        \hline
        Load Balanced           & \checkmark           & \ding{54}                   & \ding{54}            & \checkmark                          \\
        \hline
        Robustness              & Moderate             & High                        & Moderate             & High                                \\
        \hline
        Scalability             & High                 & Low                         & Moderate             & High                                \\
        \hline
    \end{tabular}
\end{table*}

\textbf{K-Connected Topologies.}
$K$-connected topologies~\cite{bonomi2019multi} with $k > f$ tolerate up to $f$ link and node failures by ensuring nodes remain connected despite such failures.
Typical examples include k-regular graphs, k-connected random graphs, k-pasted trees, k-diamonds, multipartite wheels, generalized wheels, Barabási-Albert graphs, incomplete hypercubes, and the k vertex-connected minimum spanning subgraph algorithm introduced in Wen et al.~\cite{wen2019enumerating}. The latter creates one spanning tree per sender and distributes the load across these trees to enhance robustness and efficiency by ensuring multiple disjoint paths for communication.
Our approach limits the overhead of having to maintain multiple overlay structures to a constant few, irrespective of the number of senders.

\textbf{Reliable Broadcast and Miner Extractable Value.}
Broadcast protocols that are robust to arbitrary Byzantine and even intentionally malicious behavior of nodes have been widely investigated~\cite{cachin2011introduction}.
Guerraoui et al.~\cite{DBLP:conf/wdag/GuerraouiKMPS19} guarantees consistent message delivery despite malicious nodes. Narwhal~\cite{danezis2022narwhal} is a mempool for DAG consensus protocols designed to deliver transactions more efficiently.
\lz~\cite{nasrulin2023accountable} addresses Miner Extractable Value (MEV) attacks~\cite{mazorra2022price} in which miners selectively broadcast, reorder, or suppress transactions during block creation to gain unfair advantages~\cite{daian2019flash,qin2022quantifying} (e.g., in auctions).
In addition to making miners accountable for their actions~\cite{qin2022quantifying,nasrulin2023accountable,gervais2015tampering}, MEV mitigation almost exclusively focuses on alleviating their effects~\cite{yang2022sok} (e.g., by redistributing MEV profits~\cite{flashbots-doc}).
Solutions that prevent such manipulations~\cite{zhang2020byzantine,kelkar2023themis,zarbafian2023lyra,gramoli2024aoab} require significant changes brought to the consensus layer and provide order fairness properties that differ from our dissemination fairness objective.

Our protocol builds upon and extends \lz~\cite{nasrulin2023accountable} to also address front-running attacks, while introducing dissemination fairness and improving the efficiency of dissemination.
Specifically, we force miners to disseminate transactions through randomly selected overlays to randomize when nodes learn about transactions and to prevent them form coordinating their response to such knowledge.

Table~\ref{tab:metrics_comparison} presents a comparison of dissemination approaches based on key metrics that are critical for designing fault-tolerant and scalable networks.
The dissemination mechanism describes the underlying communication strategy: Reliable broadcast~\cite{danezis2022narwhal} uses direct communication between nodes, leading to high latency for large networks because of multi-phase all-to-all message exchanges. Gossip-based dissemination~\cite{DBLP:conf/ccs/ZamaniM018,DBLP:conf/aft/RohrerT19,DBLP:conf/p2p/FreyGKM10} leverages randomized propagation, achieving moderate latency and high scalability by efficiently distributing messages while avoiding the need for complete network knowledge. Simple tree-based dissemination~\cite{neiheiser2021kauri, kogias2016enhancing} relies on fixed overlays, resulting in increasing latency for larger networks due to deeper tree structures. Its scalability is moderate because the depth of the tree scales with the number of nodes. \protocol, employing optimized robust trees, achieves low latency and high scalability by minimizing redundant paths and maintaining $f{+}1$-robust structured overlays.

Gossip-based dissemination ensures fairness. \protocol achieves the same by randomizing and optimizing dissemination trees.
Intuitively, using a larger number of overlays imply a higher bandwidth consumption, but also a lower average latency and higher dissemination fairness. Reliable broadcast and simple tree-based dissemination may suffer from imbalances, potentially leading to inconsistencies.
Reliable broadcast supports accountability, which \protocol achieves through mechanisms for detecting and tracing misbehavior, a feature absent in the other approaches.
Reliable broadcast guarantees fault tolerance by delivering to all correct nodes using a totality property. In contrast, gossip-based dissemination and simple tree-based dissemination depend on coverage and tree resilience, respectively, providing only moderate fault tolerance.
\protocol $f{+}1$-robustness ensures high fault tolerance by guaranteeing delivery even under adversarial conditions.
These attributes position \protocol as a robust, efficient, dissemination-fair, accountable, and scalable solution for large-scale, adversarial networks.

\section{System and Threat Model} 
\label{sec:system_model}

We consider a network with $n$ nodes, modeled as a labeled graph $G = (V, E)$.
Each vertex $v_i \in V$ corresponds to a node, and each edge  $e_{i,j} \in E$ represents a link between nodes $v_i$  and $v_j$.
Edges $e_{i,j}$ in the graph are labeled as $\mathit{lat}(e_{i,j})$ to capture communication latencies between nodes (here between $v_i$ and $v_j$).
We shall later maintain for each node $v_i \in G.V$ an accumulated rank (as $\mathit{rank}(v_i)$) relative to their position in the robust-tree overlay structures. Initially $0$, $\mathit{rank}(v_i)$ estimates the load of that node depending on the role it plays in the already computed overlay structures. We consider $\mathit{rank}(v_i)$ during subsequent assignments when computing the remaining overlay structures.
We assume each node is sufficiently connected so that other nodes can reach it through at least $t$ disjoint paths.

Nodes may fail in arbitrary, potentially malicious (Byzantine) ways. In contrast, links are assumed to drop messages stochastically.
We consider a fault-density threat model, defined such that within a distance of $D$ hops from any node, at most $f \le t$ nodes may fail due to accidents or cyberattacks.
For simplicity, we assume the network remains sufficiently connected despite $f$ faulty nodes in proximity.
This connectivity assumption can be extended with proof-of-connectivity messages (e.g., as proposed in Pistis~\cite{pistis_extended}).
At a broader level, fault density adheres to the classical fault tolerance threshold, allowing up to $f_{\mathit{max}}$ faulty nodes out of $n$, provided  $n \geq 3f_{\mathit{max}} + 1$.
However, the fault-density model adds as a key distinction from the classical model that no node is entirely surrounded by faulty neighbors.


\section{Overview of \protocol{}} 
\label{sec:overview}

\begin{figure*}[t]
    \begin{center}
        \includegraphics[height = 4.5cm]{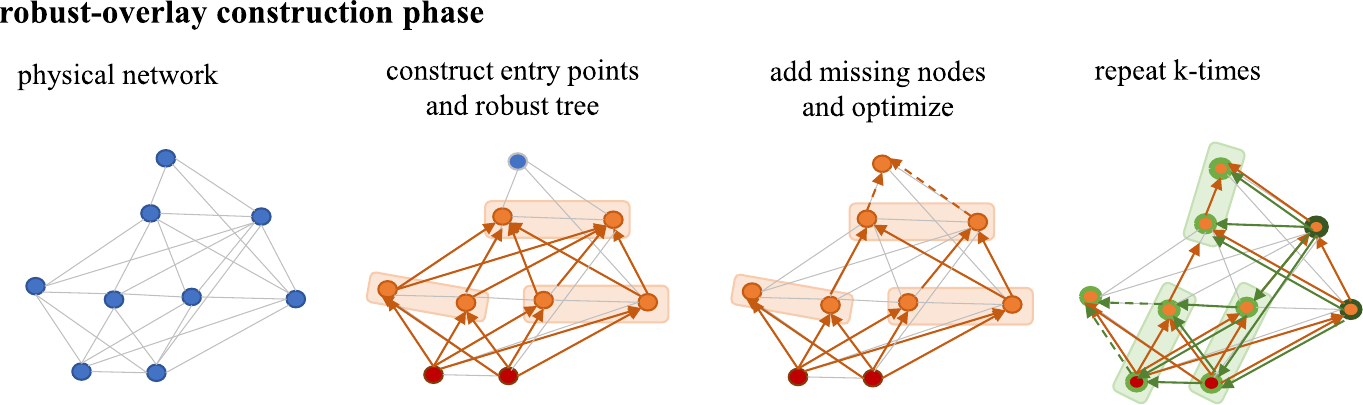}
        \includegraphics[height =4.5cm]{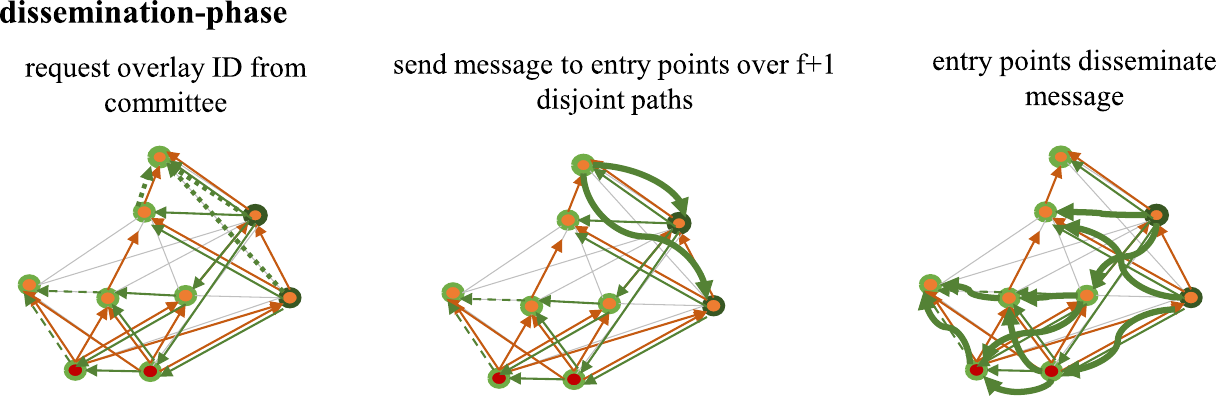}
        \caption{ Overview of \protocol{}'s two phases.
            Construction phase: $k$ robust-tree overlays are built offline and recomputed after major network changes. This involves selecting entry points, constructing robust-trees, integrating missing nodes, and optimizing for efficiency and robustness.
            Dissemination phase: A sender queries a committee to determine the overlay for message $m$ based on network state. It then forwards $m$ to entry points via $f{+}1$ disjoint paths for fault tolerance. The entry points propagate $m$ across the robust-tree overlay, ensuring reliable dissemination.
        }
        \label{fig:approach}
    \end{center}
\end{figure*}

In this paper, we are concerned with the efficient, reliable, and fair dissemination of information (transactions and blocks) through blockchain networks, while ensuring accountability to mitigate at the network layer miner-extractable-value attacks, in particular, front-running.
More precisely, we first consider the static network of permissioned blockchains and later, in Section~\ref{sec:dynamic_topologies}, the dynamic networks of permissionless blockchains. In these settings, we seek to develop a dissemination protocol that allows clients to interact with the blockchain through source nodes --- the \emph{senders} of transactions and blocks --- while tolerating locally, in the proximity of every node, up to $f$ arbitrary faulty (i.e., Byzantine) nodes. In addition, we aim to tolerate globally, in the entire network, up to $f_{\mathit{max}} = \lfloor\frac{n}{3}\rfloor$ faulty nodes.
This arbitrary behavior may, in particular, include the manipulation of transaction streams based on knowledge obtained ahead of time of other nodes.

To achieve the above, our protocol \protocol{} performs operations in two phases (see Figure~\ref{fig:approach}), which we detail in the next sections.

\noindent
\textbf{Overlay construction and optimization [offline]:}
In an offline manner, \protocol proceeds through a \emph{robust-tree overlay construction and optimization} phase, that begins by selecting $f{+}1$ entry points from the physical network G.
Using these entry points as roots, it constructs a robust-tree overlay, by identifying up to  $2^d(f{+}1)$ nodes for the new layer $d$ that are connected to all nodes of layer $d{-}1$.
Once no further layer can be constructed in this way, any missing nodes are then added to ensure robustness with $f{+}1$-connected leaves.
The structure is further optimized using simulated annealing and excess links pruned, while preserving $f{+}1$-connectivity.
This process is repeated $k$ times to construct $k$ optimized overlay structures. During optimization, the roles that each node plays in the already constructed overlays is taken into account to avoid positioning the same node at the same location in the different trees. This avoids overburdening frequently used nodes and prevents systematically favoring individual nodes across multiple overlay structures. Intuitively, using a larger $k$ value implies a higher bandwidth consumption, but also a lower average latency and higher dissemination fairness.
Once finalized, the constructed overlay structure is communicated to each node in the network. Node memorize from each overlay the entry points as well as direct predecessors and successors.

\noindent
\textbf{Dissemination [online]:}
At runtime, the dissemination of messages proceeds in three steps:
\begin{enumerate}
    \item \textbf{\emph{Randomized structure selection:}}
          The sender starts by reaching out to a committee of $3f{+}1$ nodes to ask which overlay structure it must use for transmitting its next message $m$.
          To achieve this, the sender computes a hash $H(m)$ and obtains from the committee a $2f{+}1$-threshold signature for the current transmission round $i$.
          Using this signature, the sender selects the overlay structure and forwards $m$ along with the certificate, \ie, the signature, to the $f{+}1$ entry points of this overlay structure. It does so through $f{+}1$ disjoint paths, unless of course the sender is connected directly to the overlay's entry points.

    \item \textbf{\emph{Dissemination and accountability:}}
          Entry points and intermediate nodes validate the source of the message by certifying which node sent it.
          They then forward the message to all successors in the selected overlay structure.
          These nodes also verify whether the received message follows a valid path within the overlay to detect any misbehavior.
          Nodes found to be faulty are excluded from further participation in the dissemination process.

    \item \textbf{\emph{Acknowledgment of delivery (optional):}}
          If higher-level protocols require receipt confirmations, acknowledgment messages are sent back to the sender through the same overlay structure used for dissemination.
\end{enumerate}

In each overlay structure, any node $v_i$ receives messages from at least $f{+}1$ nodes, ensuring that dissemination cannot be delayed by the up to $f$ faulty nodes near $v_i$.
Furthermore, since the overlay structure is randomly chosen by the committee (and not the sender), there is no opportunity for the sender to manipulate or choose a favorable overlay.
This mechanism ensures dissemination-fairness and accountability during the transmission process.

\begin{figure}[t]
    \begin{center}
        \includegraphics[width=\columnwidth]{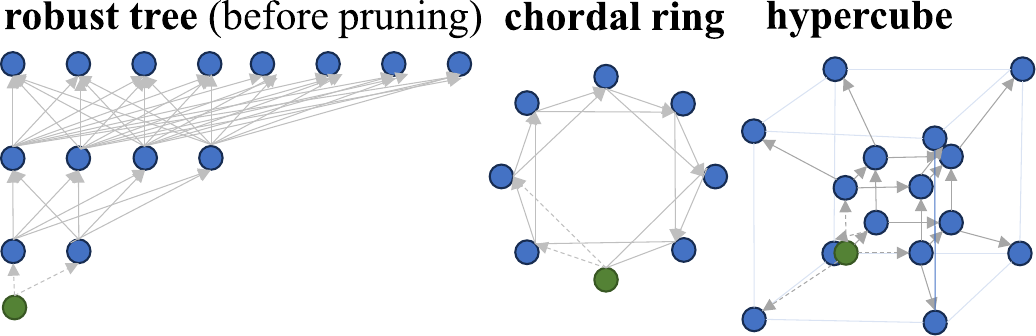}
        \includegraphics[width=\columnwidth]{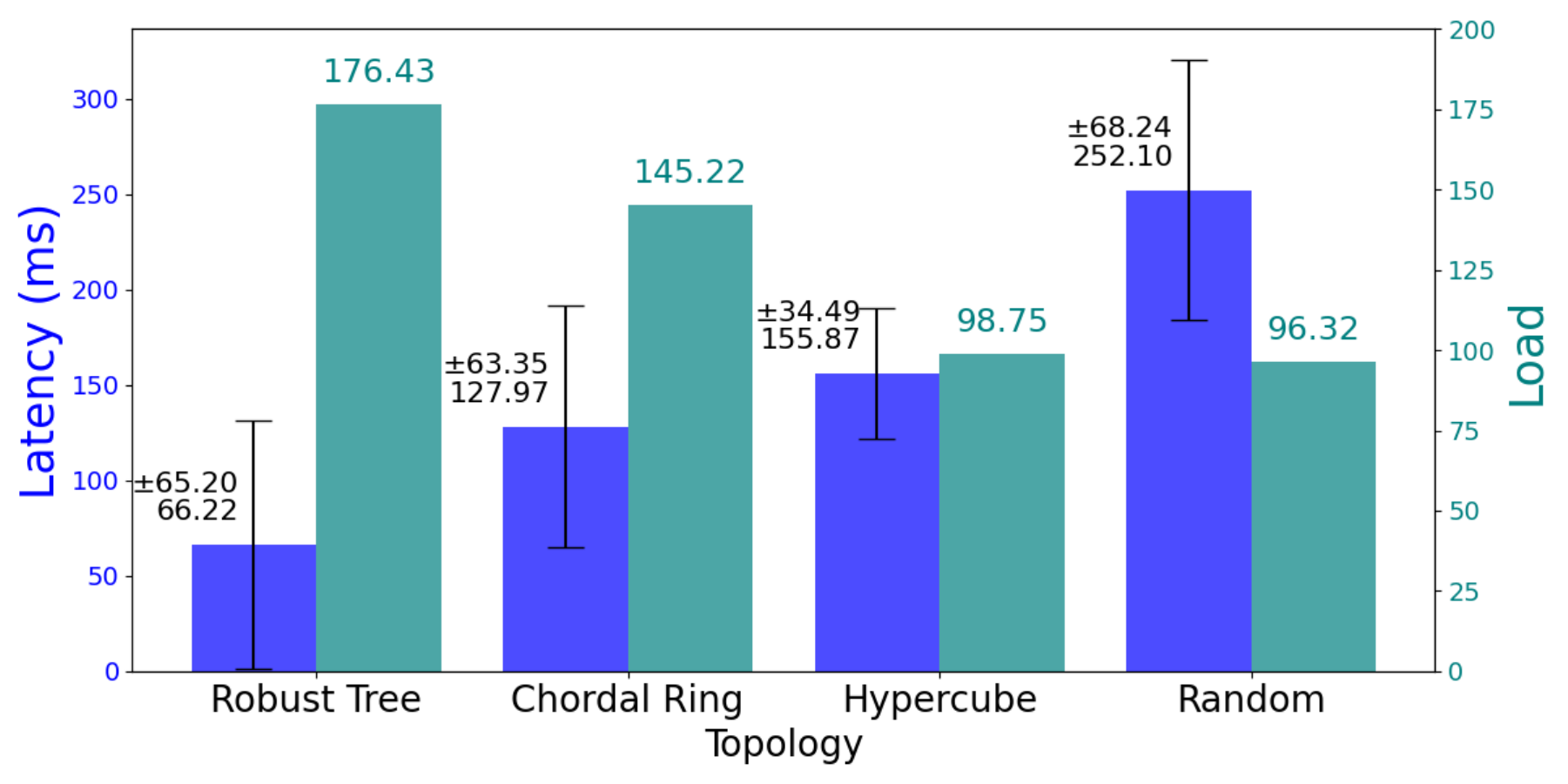}
        \caption{
            Dissemination latency and message load variance per node over a single $f{+}1$-connected instance of the considered overlays: robust trees (pre-pruning), chordal rings, hypercubes, and a random overlay ensuring at least $f{+}1$ links per node.
        }
        \label{fig:comparison_overlay}
    \end{center}
\end{figure}

\section{Overlay Structures and their Optimization} 
\label{sec:overlay}

This section motivates our selection of robust trees as overlay structure and details how we optimize them for latency.

\subsection{Overlay Structures}

Introducing overlays over the graph $G$, we aim to simultaneously achieve efficient message propagation, role-balancing, and resilience to link and node faults.
By role-balancing, we mean equally distributing dissemination responsibilities across multiple overlay structures, ensuring that no node is consistently favored against others when receiving a message, and maintaining fair participation in message propagation.
To that end, we compared $f{+}1$-connected chordal rings, hypercube, robust trees and randomly-generated overlay graphs by examining the average latency and load translated as the standard deviation of sent transactions with a constant number of nodes. The detailed setup is shown in
Section~\ref{sec:experiment_setup}.

\begin{algorithm}[t]
    \caption{CreateRobustTree}
    \label{alg:createRobustTree}

    \begin{algorithmic}[1]
        \footnotesize

        \State {\color{blue}{// Initialization}}
        \State initialize empty graphs $G^{RT}_l$ ($l \in \{1, \ldots, k\}$)
        \For{each graph $G^{RT}_l$}
        \State add $f{+}1$ entry point nodes $v^0_1, \ldots v^0_{f{+}1} \in G.V$ (nodes of rank $0$)
        \State \quad such that $v^0_i$ is among the $f{+}1$ nodes with lowest accumulated
        \State \quad rank and connecting with lowest latency to its neighbors.

        \State {\color{blue}{// Construct robust tree}}
        \For{each rank $d$ starting from $1$}
        \State select the up to $2^d(f{+}1)$ nodes $v^d_i$ in $G.V \setminus G^{RT}_l.V$ with
        \State \quad minimal acummulated rank and connection latency that are
        \State \quad connected to all nodes $v^{d-1}_j$ of the previous rank
        \State add selected nodes $v^d_i$ and their edges $e_{i,j} \in G.E$ to nodes
        \State \quad $v^{d-1}_j$ of the previous rank to the robust tree $G^{RT}_l$
        \State repeat until no further nodes fit this pattern
        \EndFor

        \State {\color{blue}{// Add missing nodes}}
        \For{each node $v_i$ remaining in $G.V \setminus G^{RT}_l.V$}
        \State add $v_i$ with $f{+}1$ edges $e_{i,j} \in G.E$ to $G^{RT}_l$, provided
        \State \quad $v^d_j \in G^{RT}_l.V$ at some rank $d$.
        \State repeat until all nodes are added
        \EndFor
        \For{each node $v^d_i \in G^{RT}_l.V$}
        \State update rank($v^d_i$) = rank($v^d_i$) + d
        \EndFor
        \State Optimize $G^{RT}_l$
        \EndFor

        \State \Return $G^{RT}_1, \ldots G^{RT_k}$

    \end{algorithmic}
\end{algorithm}

Figure~\ref{fig:comparison_overlay} illustrates the overlay structures under investigation, along with their respective latencies and the standard deviation of their load distribution. Robust trees, when considered in isolation, exhibit higher load imbalances but achieve significantly lower latency, when compared to the other overlay structures. Our approach involves disseminating messages randomly across $k$ overlay structures, the load and role imbalances in one structure can therefore be compensated by the others.
We have therefore selected robust trees as base overlay structure before pruning with simulated annealing.

Robust trees are hierarchical in nature. Nodes at a given layer (depth $d$) are connected to all nodes in the layer above them (depth $d{+}1$). However, since not all nodes in $G$ may conform to this strict pattern, the resulting robust-tree overlay, $G^{RT}$, will generally be incomplete. To address this, we augment the overlay structure by adding any remaining nodes from $G$ to $G^{RT}$, while ensuring that each such node is connected via $f{+}1$ edges either to existing nodes in $G^{RT}$ or directly within $G$. This process continues until all nodes in $G$ are integrated into $G^{RT}$.

While robust-tree overlays are highly connected, their connectivity often exceeds what is strictly necessary for reliable dissemination. Therefore, an additional optimization step is performed to prune redundant links. The result is a low-latency, $f{+}1$-connected subset of the robust tree.
Algorithm~\ref{alg:createRobustTree} outlines the pseudocode for constructing $k$ robust trees, $G^{RT}_1, \ldots, G^{RT}_k$, from the physical network $G$: Starting with an empty set of graphs, the algorithm selects $f{+}1$ nodes with the lowest accumulated rank and minimal latency to their neighbors as the entry points and assigns them to rank $d=0$ (Lines 3--6).
The tree, now initiated with $f{+}1$ roots, is extended iteratively. For each rank, nodes are added that are directly connected to all nodes in the previous rank (depth $d-1$) and have not yet been included in $G^{RT}_l$ (Lines 8--15).
Nodes are selected based on their minimal accumulated rank to balance the roles they will obtain and based on their connection latency to neighboring nodes. At each rank, the goal is to double the number of included nodes, such that at depth $d$ there are $2^d(f{+}1)$ nodes.
If doubling is no longer feasible, remaining nodes are added without enforcing the hierarchical tree structure, ensuring they connect to at least $f{+}1$ edges of existing nodes in $G^{RT}_l$ (Lines 17--21).
Before passing the constructed tree to the optimization step (Line 25) and moving to the next tree, the accumulated rank of nodes is updated to reflect their distance from the root in the current tree (Lines 22--24).
This ensures a balanced distribution of nodes across the $k$ robust trees.
The described methodology ensures the creation of $k$ robust, low-latency overlay structures optimized for scalability, fault tolerance, and efficiency.

\subsection{Overlay Latency Minimization and Role-Balancing}

\begin{algorithm}[t]
    \caption{Simulated Annealing for Multiple Overlay Optimization}
    \label{alg:simulated}
    \begin{algorithmic}[1]

        \State \textbf{Input:} Network graph $G$ and overlay $O_{\text{current}} = G^{RT}_l$, initial temperature $T_{\text{initial}}$, minimum temperature $T_{\text{min}}$, cooling rate $\alpha$
        \State \textbf{Output:} Optimized overlays $O_{\text{best}}$

        \State {\color{blue}{// Step 1: Initialize Overlay Optimization}}

        \State $O_{\text{best}} \gets O_{\text{current}}$
        \State $T \gets T_{\text{initial}}$

        \State {\color{blue}{// Step 2: Perform Simulated Annealing}}
        \While{$T > T_{\text{min}}$}
        \State // Generate neighbor overlay:
        \State $O_{\text{neighbor}} \gets \textbf{GenerateNeighbor}(O_{\text{current}}, f)$
        \State // Evaluate objective values:
        \State $O_{\text{neighbor}} \gets \textbf{ObjectiveFunction}(O_{\text{neighbor}})$
        \State $O_{\text{current}} \gets \textbf{ObjectiveFunction}(O_{\text{current}})$

        \If{$O_{\text{neighbor}} < O_{\text{current}}$\\
            \qquad\qquad \textbf{or} $e^{-(O_{\text{neighbor}} - O_{\text{current}})/T} > \text{rand(0, 1)}$}
        \State Accept neighbor overlay: $O_{\text{current}} \gets O_{\text{neighbor}}$
        \If{$o_{\text{neighbor}} < \textbf{ObjectiveFunction}(O_{\text{best}})$}
        \State Update best overlay: $O_{\text{best}} \gets O_{\text{neighbor}}$
        \EndIf
        \EndIf

        \State Decrease temperature: $T \gets T \times \alpha$
        \EndWhile

        \State \textbf{Return} $O_{\text{best}}$

    \end{algorithmic}
\end{algorithm}

The goal of this step is to remove redundant network links while maintaining $f{+}1$-connectivity, and to optimize the overlay structures for both latency and role distribution.
Indeed, a node positioned near the roots of multiple tree overlays would be advantaged, as it would tend to receive messages earlier on average compared to its neighboring nodes.
Additionally, in a tree structure, nodes closer to the root typically bear higher loads than those near the leaves.
To address this imbalance, we introduce a penalty system based on the accumulated ranks of nodes across previously constructed overlay structures.
The accumulated rank reflects how often a node has been placed near the leaves in past overlays.
Nodes with higher accumulated ranks are more likely to have been less loaded, making them preferable candidates for near-root positions in the current overlay structure.

\textbf{Latency Optimization.}
In this work, we employ simulated annealing as optimization algorithm, although our approach is agnostic to the specific optimization algorithm employed.
Simulated annealing iteratively improves the overlay structure by exploring configurations that minimize latency and complexity while factoring in rank penalties.
The algorithm starts with the robust tree $G^{RT}_l$, which initially relies on a heuristic for node inclusion, and incrementally explores modifications to improve the overlay. These modifications may include adjusting node connections or rerouting specific links. At each iteration, the algorithm evaluates whether a proposed change improves the overall objective function. Changes are accepted probabilistically, based on the current system temperature, \ie, how far it is from the overall objective, which decreases over time. This decreasing temperature enables broad exploration in the early stages and gradual convergence toward an optimal or near-optimal solution. The pseudocode for the simulated annealing algorithm is provided in Algorithm~\ref{alg:simulated}.

\textbf{Balancing roles using rank penalty.}
In addition to optimizing for latency, we incorporate a rank penalty to discourage the repeated selection of nodes that have already been placed as low-rank nodes (near the root) in other overlay structures.
Nodes closer to the root experience earlier message delivery and higher loads. It is therefore preferable to position them closer to the leaves in subsequent overlay structures. To achieve this, we define the objective function as:

\begin{equation}
    \label{eq:objective}
    \begin{array}{rl}
        \mathit{objective\_value} = & \mathit{num\_edges} + \mathit{avg\_latency}       \\
                                    & + \mathit{connectivity\_penalty}                  \\
                                    & + \mathit{path\_penalty} + \mathit{rank\_penalty}
    \end{array}
\end{equation}

where $\mathit{num\_edges} = |E^{RT}_l|$ (from $G^{RT}_l = (V, E^{RT}_l) \subseteq G$), $\mathit{avg\_latency}$ the sum of all shorted path latencies in $G^{RT}_l$ divided by $|V| = n$.
We penalize nodes with less than $f{+}1$ successors ($\mathit{connectivity\_penalty}$), nodes that are not reachable from the source ($\mathit{path\_penalty}$) and discourage the use of nodes with low accumulated rank ($\mathit{rank\_penalty}$).

Algorithm~\ref{alg:generateneighbor_optimized} illustrates the process for generating a neighboring solution from the current robust-tree overlay configuration.
The function begins by randomly adding or removing edges between nodes in consecutive layers of the overlay.
This random adjustment enables exploration of alternative configurations.
To ensure resilience, the algorithm then verifies that each node (except leaves) retains at least $f{+}1$ outgoing connections.
Additionally, the function considers the depth and accumulated rank of nodes, reassigning connections where necessary to balance the role and minimize the rank penalty.
By simultaneously optimizing connectivity and latency while reducing the rank penalty, the function generates valid neighbor solutions that preserve robustness and enable a dissemination-fair framework.
In theory, highly heterogeneous latency distributions might prevent our optimization procedure from obtaining a high level of dissemination fairness among nodes. However, this effect would be attenuated as the system size increases, and could also be specifically mitigated by modifying our simulated annealing optimization procedure.

\begin{algorithm}[t]
    \caption{GenerateNeighbor}
    \label{alg:generateneighbor_optimized}
    \begin{algorithmic}[1]

        \State \textbf{Input:} Current robust-tree overlay $G^{RT}_l$, fault tolerance $f$, accumulated rank $rank(v)$
        \State \textbf{Output:} Neighbor robust-tree overlay $G^{RT}_{\mathit{neighbor},l}$

        \Function{GenerateNeighbor}{$G^{RT}_l, f, rank(v)$}
        \State Create a copy of $G^{RT}_l$ named $G^{RT}_{\mathit{neighbor},l}$
        \State $layers \gets$ list of node layers in $G^{RT}_l$

        \State {\color{blue}{// Step 1: Randomly Add or Remove an Edge}}
        \If{random number $< 0.5$ \\
            \quad\qquad \textbf{and} $G^{RT}_{\mathit{neighbor},l}$ has more than 0 edges}
        \State Choose a random edge between consecutive layers
        \State \quad and remove it
        \Else
        \State Choose a random edge between consecutive layers
        \State \quad and add it
        \EndIf

        \State {\color{blue}{// Step 2: Ensure $f{+}1$-Connectivity}}
        \For{each layer in $layers$}
        \If{layer $< \max(layer\_keys)$}
        \For{each node in $layers[layer]$}
        \While{node has $< f + 1$ successors}
        \State Add an edge from node to a random
        \State \quad node in the next layer, minimizing:
        \State \qquad $\mathit{latency}(node, next\_node)$
        \EndWhile
        \EndFor
        \EndIf
        \EndFor

        \State {\color{blue}{// Step 3: Adjust for Rank Penalty}}
        \For{each node in $G^{RT}_{\mathit{neighbor},l}$}
        \If{depth of node $i$ is too close to root and \\\qquad\qquad has extra edges}
        \State Attempt to reassign some connections to
        \State \quad nodes with higher accumulated rank $\mathit{rank}(v)$
        \State \quad and farther depth
        \EndIf
        \EndFor

        \State {\color{blue}{// Step 4: Validate and Return Optimized Neighbor}}
        \If{$\text{ObjectiveFunction}(G^{RT}_{\mathit{neighbor},l}, rank)$\\
            \qquad\quad $ < \text{ObjectiveFunction}(G^{RT}_l, rank)$\\\quad}
        \State \Return $G^{RT}_{\mathit{neighbor},l}$
        \Else
        \State \Return $G^{RT}_l$ \Comment{Discard if no improvement}
        \EndIf
        \EndFunction

    \end{algorithmic}
\end{algorithm}

\begin{algorithm}[t]
    \caption{Threshold Random Seed (TRS) Generation via Reliable Broadcast}
    \label{alg:trs_generation_rbroadcast}
    \begin{algorithmic}[1]

        \State \textbf{Input:} Committee of $3f{+}1$ nodes, sequence number $i$, message hash $H(m)$, source node $N_{\text{source}}$
        \State \textbf{Output:} Threshold Random Seed $\varphi(i, H(m))$

        \State {\color{blue}{// Step 1: Broadcast by Source Node}}
        \State $N_{\text{source}}$ sends $(i, H(m))$ to all committee members

        \State {\color{blue}{// Step 2: Reliable Broadcast Among Committee Members}}
        \For{each committee member $n_j$}
        \State On receiving $(i, H(m))$:
        \begin{itemize}
            \item Send an ``Echo'' message containing $(i, H(m))$ to all other committee members.
        \end{itemize}
        \State On receiving \( 2f{+}1 \) ``Echo'' messages or \( f{+}1 \) ``Ready'' messages:
        \begin{itemize}
            \item Send a ``Ready'' message containing $(i, H(m))$ to all other committee members.
        \end{itemize}
        \State On receiving \( 2f{+}1 \) ``Ready'' messages:
        \begin{itemize}
            \item Deliver the message $(i, H(m))$.
        \end{itemize}
        \EndFor

        \State {\color{blue}{// Step 3: Partial Signature Generation and Collection}}
        \For{each committee member $n_j$ who delivered $(i, H(m))$}
        \State Compute partial signature: $\sigma_j = \text{Sign}(n_j, (i, H(m)))$
        \State Send $\sigma_j$ to $N_{\text{source}}$
        \EndFor

        \State {\color{blue}{// Step 4: Combine Partial Signatures at Source Node}}
        \State Source node $N_{\text{source}}$ collects all received $\sigma_j$ into $\text{partial\_signatures}$

        \If{$|\text{partial\_signatures}| \geq 2f{+}1$}
        \State Combine partial signatures: $\varphi(i, H(m)) = \text{CombineSignatures}(\text{partial\_signatures})$
        \State \textbf{Return} $\varphi(i, H(m))$
        \EndIf

    \end{algorithmic}
\end{algorithm}

\section{Random, Accountable and Fair Dissemination}

\label{sec:account-fair}

To prevent successful front-running attacks, \protocol employs distributed and resilient mechanisms that enable fair, randomized and accountable overlay selection and message dissemination.
Together, these mechanisms strive towards
dissemination-fairness,
ensuring that no correct node is systematically left behind in the delivery of messages and no adversary can exploit timing advantages to reorder messages or front-run their dissemination.
\protocol's accountability ensures that any deviations from the prescribed dissemination path can be detected and flagged as protocol violations.
In a nutshell, \protocol's dissemination follows three main steps:
(i)~a committee of $3f{+}1$ nodes generates a TRS for each message $m$,
(ii)~the TRS is used to select a random overlay structure for message dissemination,
(iii)~nodes verify the legitimacy of the message they receive and the senders of this message according to the known overlay structures before relaying it to their successors.

\subsection{Threshold Random Seed (TRS) Generation}
\label{sec:trs-gen}
To achieve a fair and unbiased overlay selection, \protocol generates a
threshold random seed per message, which is then used to determine uniformly at random the overlay structure to be used for dissemination.
This process builds on a reliable broadcast algorithm~\cite{bracha1987asynchronous}, ensuring consistency and preventing any node from unilaterally affecting the random seed used later in the protocol.

When a node intends to broadcast a message $m$, it first sends the hash $H(m)$ along with a monotonically increasing sequence number $i$ to a committee of $3f{+}1$ nodes.
The sequence number $i$ ensures that messages are processed in ascending order without duplication.
Committee members echo $(i, H(m))$ to one another, and each member dispatches a ``Ready'' message upon receiving either $2f{+}1$ ``Echo'' messages or $f{+}1$ ``Ready'' messages.
Upon reception, each member creates a partial threshold signature and sends it back to the source node.
The source collects $2f{+}1$ partial signatures, combines them, and obtains a final threshold signature $\varphi(i, H(m)$ that effectively acts as a random seed for the transaction $m$ in round $i$.
Its distributed and f-tolerant generation process guarantees that no single adversarial node (or small group) can skew the seed to benefit front-running or any other manipulation.
Moreover, because each message is bound to a unique sequence number, malicious committee members cannot successfully modify it to front run, reorder or skip messages without detection during seed generation.

\subsection{Randomized and Verifiable Overlay Selection}
\label{sec:overlay-trs}

Once the TRS $\varphi(i, H(m))$ is generated, the source node computes \emph{overlay = seed mod $k$}, where $seed = \varphi(i, H(m))$, and $k$ is the total number of available overlays.
This step ensures a randomized but verifiable selection of overlay for every message, thereby hindering adversaries from systematically front-running.
Consequently, all nodes can independently verify the threshold signature and the selected overlay, preventing dishonest senders from unilaterally choosing favorable routes.
Additionally, no node can predict or predetermine which overlay will be used for $m$’s propagation, ensuring an even playing field for all correct nodes.

\subsection{Overlay Encoding and Accountable Dissemination}
\label{sec:encoding-dissemination}
Alongside randomized overlay selection, \protocol employs \emph{accountable} dissemination to guarantee that messages follow the correct path and that nodes can detect and expose any front-running attempts or deviations.

Before dissemination begins (or when overlays are re-optimized for a permissionless setting), \protocol distributes to every node a signed description of the $k$ \emph{robust-tree} overlays used.
Algorithm~\ref{alg:robust_tree_encoding} provides the details of this \emph{Robust Tree Encoding}.
Each node learns its predecessors and successors within each overlay, as well as the designated entry points.
These entry points enable nodes to forward or receive messages in accordance with the chosen overlay.

A node receiving a message $m$ directly verifies the message's and sender's legitimacy by checking
(i)~the threshold signature $\varphi(i, H(m))$,
(ii)~the sequence number $i$, and
(iii)~that the sender is a valid predecessor in the overlay.
If any check fails, the recipient flags the message as a protocol violation.
This auditing step holds each node accountable for adhering to the assigned route.

Through the TRS generation process, the committee binds the message $m$ that a node wishes to send as its $i^{th}$ message to the random seed they generate, ensuring continuity of sequence numbers.
This mechanism prevents dissemination omissions by requiring a node to first transmit any missing messages for skipped sequence numbers before sending the current message as its $i^{th}$ message.
Nodes cannot omit or reorder messages without detection, since skipping a sequence or forging a threshold signature is easily caught by other nodes.
By enforcing this requirement, the protocol ensures compliance and prevents attempts to bypass the sequence ordering and front run.
This ensures tamper-proof evidence of each transmission path, enabling the system to isolate and exclude deviant nodes from future communications.

Because decisions about routing and overlay choice depend on threshold signatures
rather than a centralized authority, \protocol remains decentralized.
Nodes themselves monitor compliance through local checks and cross-verification, ensuring dissemination-fairness and resilience against both targeted node overloads and front-running attempts.

\begin{algorithm}[t]
    \caption{Robust Tree Encoding}
    \label{alg:robust_tree_encoding}
    \begin{algorithmic}[1]

        \State \textbf{Input:} Committee of $3f{+}1$ nodes, source node $N_{\text{source}}$
        \State \textbf{Output:} Encoded robust tree $\mathcal{T}_{\text{encoded}}$ and threshold signature $\sigma_{\mathcal{T}}$

        \State {\color{blue}{// Step 1: Robust Tree Computation}}
        \For{each committee member $n_j$ \textbf{in parallel}}
        \State Compute robust tree $\mathcal{T}_j$
        \State Encode robust tree: $\mathcal{T}_{\text{encoded}} = \text{EncodeTree}(\mathcal{T}_j)$
        \State Generate partial signature: $\sigma_j = \text{Sign}(n_j, \mathcal{T}_{\text{encoded}})$
        \State Send $\mathcal{T}_{\text{encoded}}$ and $\sigma_j$ to $N_{\text{source}}$
        \EndFor

        \State {\color{blue}{// Step 2: Threshold Signature Generation at Source Node}}
        \State Source node $N_{\text{source}}$ collects received $\sigma_j$ into $\text{partial\_signatures}$

        \If{$|\text{partial\_signatures}| \geq 2f{+}1$}
        \State Combine partial signatures: $\sigma_{\mathcal{T}} = \text{CombineSignatures}(\text{partial\_signatures})$
        \State \textbf{Return} $\mathcal{T}_{\text{encoded}}, \sigma_{\mathcal{T}}$
        \EndIf

    \end{algorithmic}
\end{algorithm}




\section{Discussion}
\label{sec:dynamic_topologies}

\subsection{Fault Density Assumption}

The fault-density assumption is fundamental for the use of robust overlays in \protocol{}. It ensures that within a radius of $D$ hops, no node is surrounded by more than $f$ faulty neighbors.
This assumption is particularly practical for permissioned systems, where the network topology can be controlled or monitored, at least partially.
However, in permissionless environments, such as public blockchains, this assumption may not always hold due to factors like malicious node clustering or dynamic network churn.

When fault density is violated, \protocol{} may encounter degraded performance or incomplete message dissemination.
Specifically, malicious nodes clustered around a target node may try to isolate and prevent it from reliably receiving or forwarding transactions.
Similarly, fault-density violations can lead to disproportionately high workloads at healthy nodes, reducing efficiency and dissemination fairness.

To address these scenarios, we incorporate gossip as robust fallback mechanism, activated after a delay $T$ to give \protocol{}'s overlays enough time to disseminate, and leverage the reconciliation mechanism of mempools built on top of \protocol{}.
Gossip operates in the background to ensure eventual consistency and delivery of transactions.
If a node fails to receive a certain messages due to faults or adversarial interference, the mempool's reconciliation protocol identifies and retransmits the missing transactions through gossip to ensure reliable dissemination despite initial disruptions, even in scenarios where fault-density assumptions cease to hold.

\subsection{Dynamic Network Topologies}

\protocol{} can be directly integrated into blockchains that rely on epoch-based memberships (e.g., Ethereum~\cite{buterin2013ethereum}): overlays would simply have to be recomputed at the beginning of an epoch.
Extending \protocol{} to other permissionless blockchains with dynamic network topologies is achieved by integrating a Byzantine-fault-tolerant gossip-based peer-sampling mechanism, such as RAPTEE and SecureCyclon~\cite{pigaglio2022raptee, antonov2023securecyclon}.
This mechanism allows each node to maintain a local, partial view of the network membership, which is periodically refreshed by exchanging information with other nodes. This dynamic update process ensures that even in environments with high churn or malicious nodes attempting to over-represent themselves, every node remains $f{+}1$ connected and has an up-to-date understanding of the network topology.

When nodes join or leave the network, \protocol{} adapts by updating the overlay structures. Newly joined nodes are integrated into all overlay structures with at least $f{+}1$ connectivity, maintaining the system's fault tolerance. Departing nodes are removed from the overlays while ensuring the necessary level of connectivity for the remaining nodes.

These dynamic updates inevitably cause \protocol{} to diverge from its optimized structure. To address this, \protocol{} periodically re-executes the steps of the offline overlay construction phase to create new optimized overlays that reflect the current topology during the next epoch.
This reconstruction process occurs in parallel and in the background, ensuring that the existing overlay remains operational until the updated structure is disseminated and confirmed.
The reconstruction process can either be offloaded to a trusted data center operated by the blockchain provider or executed within the blockchain network itself in a distributed manner. In the latter case, \protocol{} mitigates attempts by faulty nodes to manipulate the generated structure by validating each step of the construction with $f{+}1$ neighboring nodes. Additionally, the committee ensures deterministic construction by generating a random seed for use in the pseudo-random optimization steps performed by simulated annealing.

A special case arises when an entry-point node in an overlay structure leaves the network. In this scenario, a new entry point must be elected and disseminated to the network. During this transition, the overlay structure may temporarily become unusable if the remaining $f$ entry-point nodes are faulty. However, the global view maintained through the gossip mechanism allows \protocol{} to continue efficiently disseminating transactions and blocks across the network. This ensures low latency, fairness, and resilience against adversarial attacks, such as front-running.



\section{Performance Evaluation}  
\label{sec:evaluation}
We address the following questions:
\begin{itemize}
    \item How does \protocol{} compare to the state-of-the-art protocols, \lz, Mercury, and Narwhal, in terms of latency under the same network conditions?
    \item What is the bandwidth overhead associated with each protocol, and how does it impact transaction propagation efficiency?
    \item How does each protocol perform in the presence of Byzantine nodes, and which protocol demonstrates the highest level of robustness and fairness?
\end{itemize}

We conducted simulations to compare the protocols in controlled environments, evaluating key metrics such as latency, and bandwidth usage. Additionally, we analyzed their performance under various attack scenarios, including front-running and censorship, to assess their resilience and ability to maintain fairness and robustness. This thorough evaluation allows us to identify the strengths and weaknesses of each protocol across diverse network conditions.

\subsection{Experimental Setup} 
\label{sec:experiment_setup}
We evaluated our protocol and the baselines on a server equipped with Intel Xeon Gold 6240 CPUs (36 physical cores, 72 logical CPUs) running at 2.6 GHz (boosting up to 3.9 GHz) and 128 GiB of RAM, interconnected via Gigabit Ethernet. The experiment parameters included $N=10,000$ nodes, a transaction size of 250 bytes, repeating each experiment 10 times and reporting average results.

All protocols were thoroughly developed entirely in Python, based on a single, common P2P simulation framework. The Mercury implementation used was the complete version, including the early outburst and clustering features, with parameters set to: $K = 8$ clusters, $D_{cluster} = 4$ inner-cluster peers, and $D_{max} = 8$ maximum peers. Both Narwhal and \lz{} use connected topology.
In \protocol we use optimized robust trees with $f=1$ and tree depth $D=14$ following the number of nodes used for the experiments $N$, and considering 10 different overlays. Under those settings, computing the overlays took less than 15\,s.

To simulate realistic network conditions, we used a latency dataset derived from multiple sources, including CAIDA~\cite{caida_topology}, RIPE Atlas~\cite{ripe_atlas}, and major cloud providers such as AWS~\cite{aws_latency} and Azure~\cite{azure_latency}, alongside empirical blockchain latency data (Ethereum~\cite{ethereum_latency}). Latencies were modeled using inverse gamma distributions for intra-regional communication and normal distributions for inter-regional latencies, covering key regions such as New York, Singapore, Frankfurt, Sydney, Tokyo, Ireland, Ohio, California, and London. Specifically:

\begin{itemize}
    \item \textbf{Intra-regional communication} (e.g., nodes within Frankfurt) followed an inverse gamma distribution with shape $\alpha = 2.5$ and scale $\beta = 14$, resulting in a mean latency of $7$ ms and a variance of $2$ ms.
    \item \textbf{Inter-regional communication} (e.g., Frankfurt to New York) followed a normal distribution with mean $\mu = 90$ ms and variance $\sigma^2 = 20$ ms.
\end{itemize}

\subsection{Comparative Analysis} 
\label{sec:Comparison}

\begin{figure}[t]
    \begin{center}
        \begin{subfigure}{.45\textwidth}
            \includegraphics[width=\columnwidth]{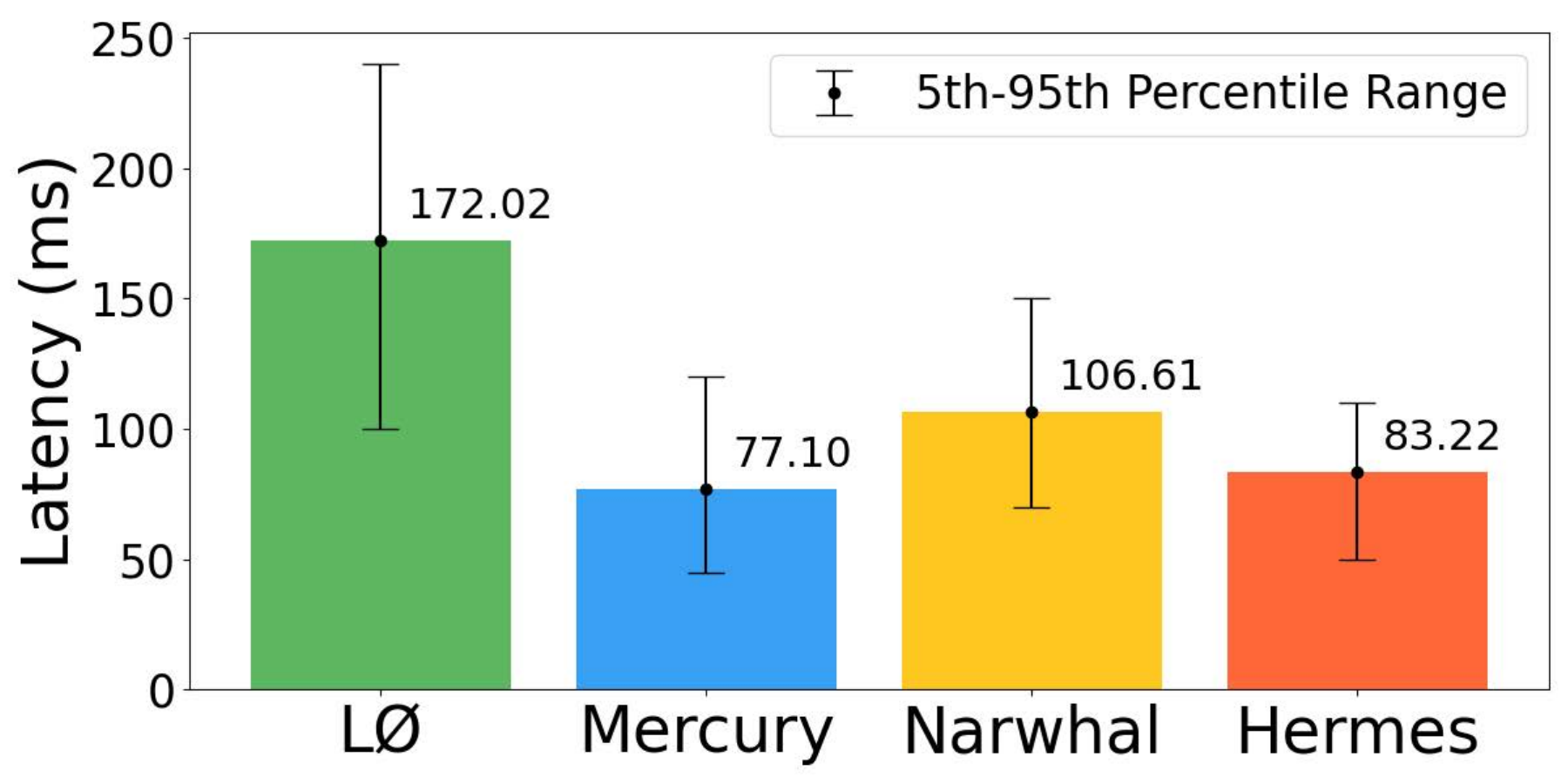}
            \caption{Average Latency}
            \label{fig:latency}
        \end{subfigure} \hfill
        \begin{subfigure}{.45\textwidth}
            \includegraphics[width=\columnwidth]{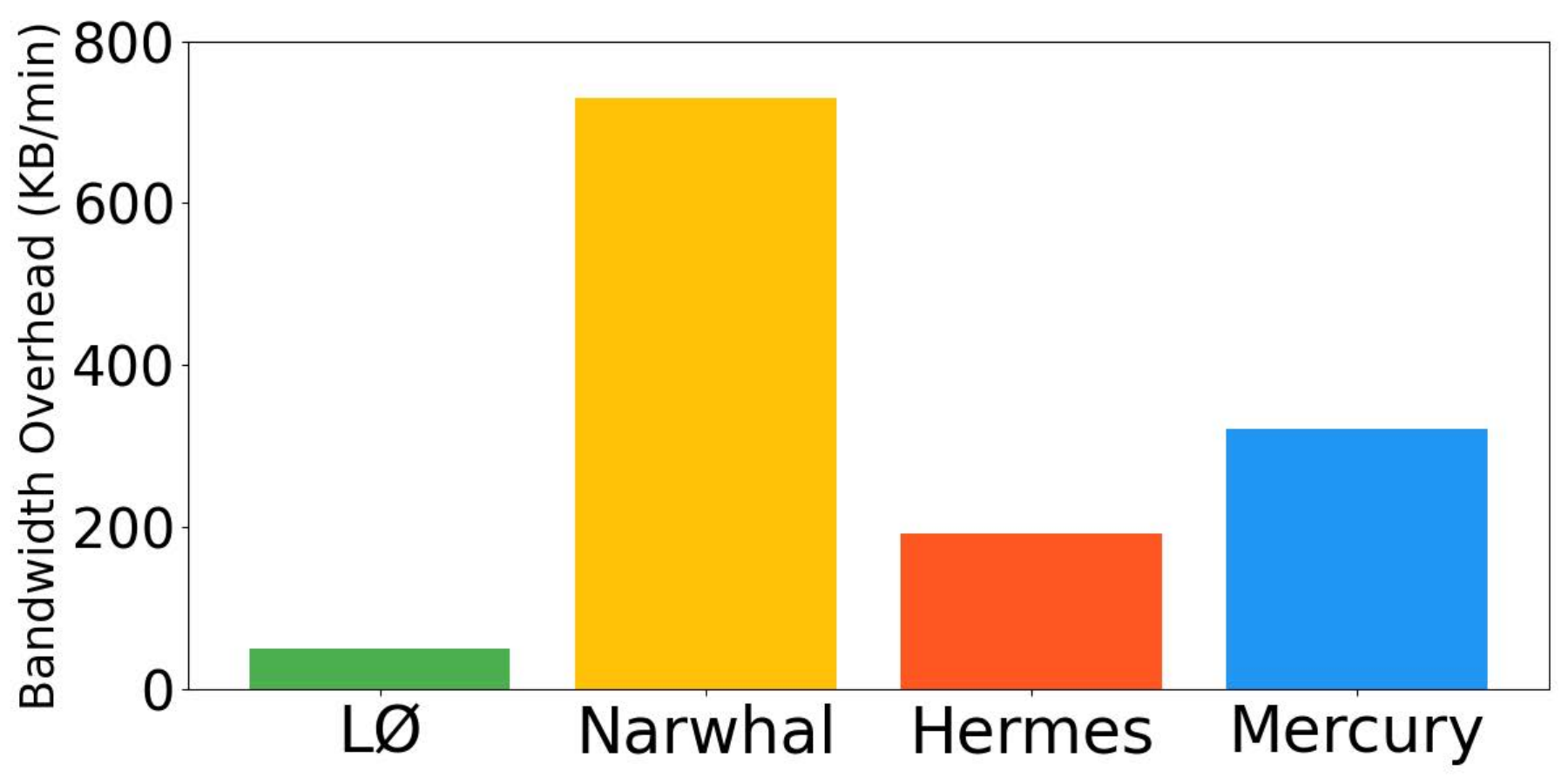}
            \caption{Bandwidth overhead}
            \label{fig:overhead}
        \end{subfigure}
    \end{center}
    \caption{Achieved performance and comparison to \lz{}, Mercury and Narwhal.}
    \label{fig:performance}
\end{figure}

Our comparative analysis examines the performance of four baselines: \protocol{} (our protocol), Narwhal, \lz{}, and Mercury. Narwhal~\cite{danezis2022narwhal} is a mempool protocol designed for enhancing consensus performance and relies on standard broadcasting. \lz{}~\cite{nasrulin2023accountable} is a lightweight mempool reconciliation protocol that optimizes transaction propagation with minimal bandwidth. Mercury~\cite{zhou2023mercury}, a clustering-based propagation protocol, aims to optimize transaction dissemination latency. We evaluated these protocols in terms of their average latency (measuring transaction propagation times), latency distribution (assessing consistency of propagation), scalability (evaluating how well the protocol handles larger networks), and overhead (measuring computational and communication costs). We also conducted a security analysis to provide insights into each protocol’s strengths and weaknesses.\\

\subsection{Transaction Latency}
We evaluate the average transaction dissemination latency and its variability, as shown in Figure~\ref{fig:latency}. Average latency measures the typical delay, while latency variability is represented by the range between the 5th and 95th percentiles that captures the spread of latency values. Together, these metrics provide a comprehensive view of each protocol's efficiency and consistency in transaction propagation.
Mercury achieves the lowest average latency of $77.10$\,ms and a relatively narrow latency range thanks to its clustering-based dissemination approach.
However, slight variability arises from inter-cluster communication overhead.
\protocol{} achieves a good latency of 83.22 ms demonstrating slightly higher latency than Mercury. However, \protocol{} achieves lower variability by generating dissemination paths that are carefully optimized for balanced load and minimized latency variability using simulated annealing. The additional overhead from TRS and tree encoding slightly increases the average latency but does not contribute significantly to variability.
Narwhal, with an average latency of 106.61 ms, demonstrates moderate transaction speeds, but its broader latency range indicates less consistency in propagation times due to coordination dependencies between nodes.
\lz{} exhibits the highest latency at 172.02 ms, with the widest latency range, reflecting the inefficiency inherent in traditional gossip mechanisms, where propagation relies on repeated message exchanges among nodes.

\begin{figure*}[t]
    \centering
    \includegraphics[width=.85\linewidth]{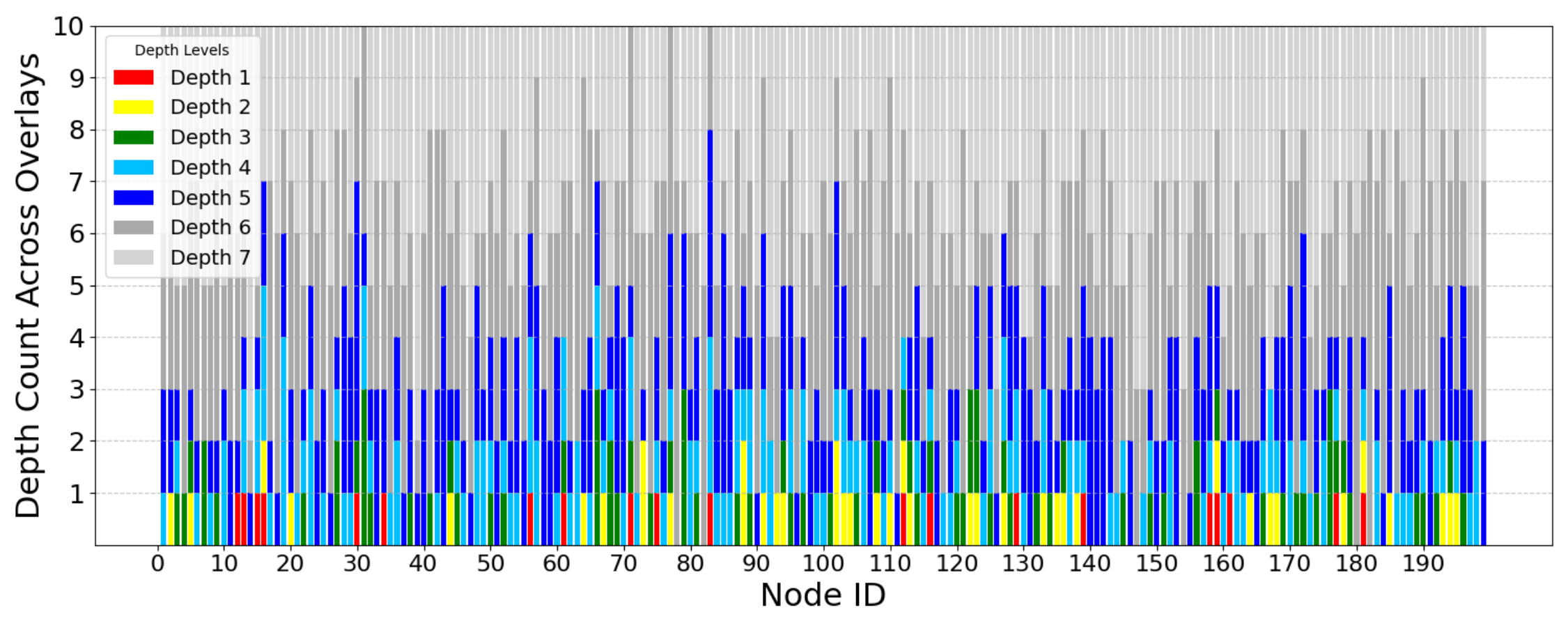}
    \caption{
        Distribution of roles for each of the 200 nodes evaluated across all 10 generated overlay structures.
        Shown are the number of overlay structures in which each node held a given rank. For example, node 30 is the entry point in one overlay, ranked 3 in another, ranked 5 in five more overlays, ranked 6 in two, and ranked 7 in one.
    }
    \label{fig:depth}
\end{figure*}

\begin{figure}[t]
    \begin{center}
        \begin{subfigure}{.45\textwidth}
            \includegraphics[width=\columnwidth]{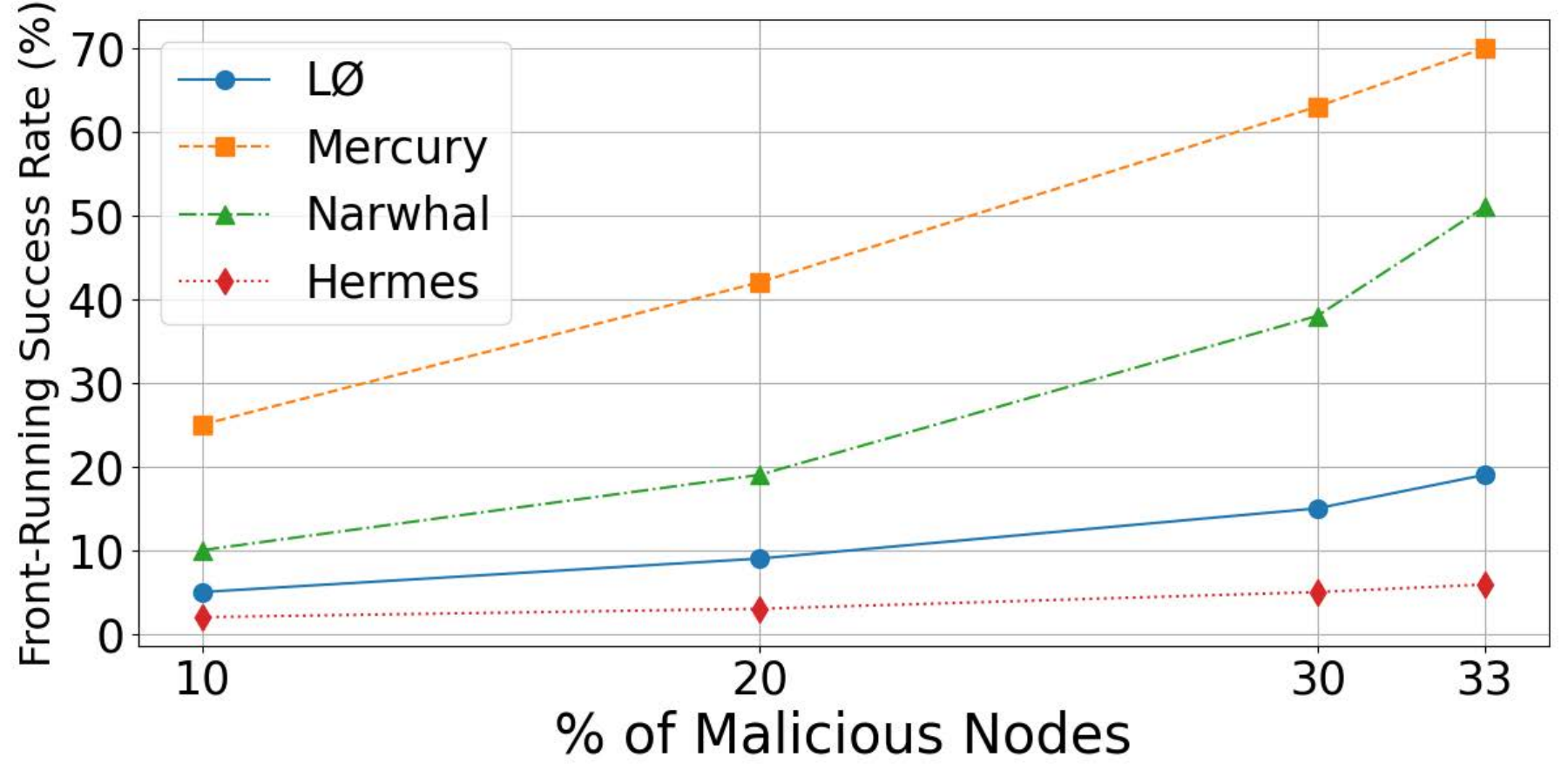}
            \caption{Front-running success rate depending on the proportion (\%) of malicious nodes}
            \label{fig:front-run}
        \end{subfigure}
        \hfill
        \begin{subfigure}{.45\textwidth}
            \includegraphics[width=\columnwidth]{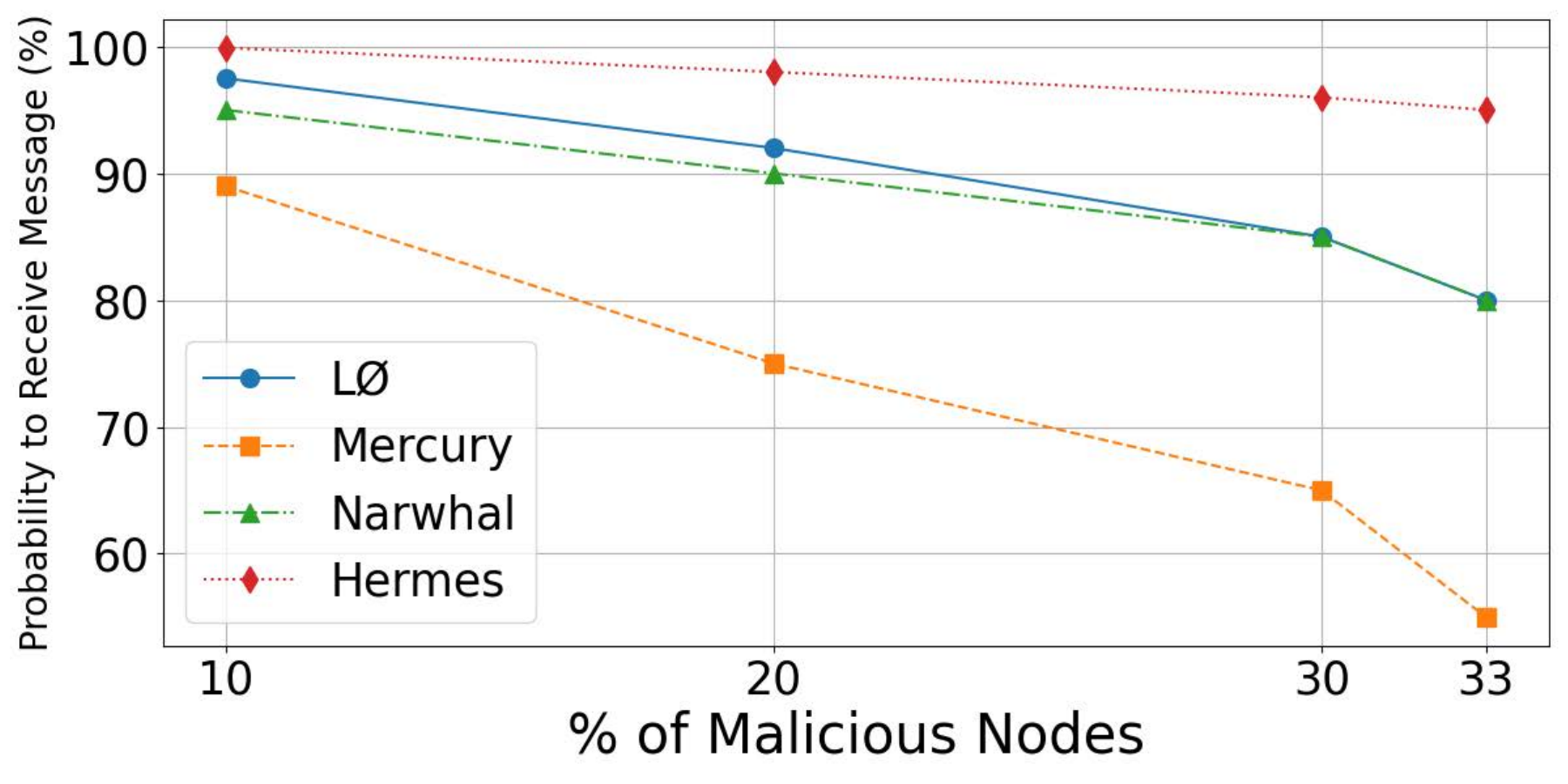}
            \caption{Probability for a message to be received. }
            \label{fig:robustness}
        \end{subfigure}
        \caption{Front-running resistance and robustness}
        \label{fig:fair_robust}
    \end{center}
\end{figure}

\subsection{Bandwidth Consumption}
Figure~\ref{fig:overhead} shows the differences in bandwidth overhead among \lz{}, Mercury, Narwhal, and \protocol{}, highlighting how each protocol balances efficiency and performance. For simplicity, we collect results for $N=200$ nodes.
\lz{} demonstrates the lowest bandwidth overhead, with only 50 KB/min, attributed to its highly efficient mempool reconciliation mechanism, making it an excellent choice for resource-constrained environments.
\protocol{}, designed as an enhanced version of \lz, introduces mechanisms such as the Threshold Random Seed (TRS) and a compact tree encoding, which prioritize fairness and latency optimization. The measured bandwidth overhead for \protocol{} is 192 KB/min, assuming frequent dissemination of the encoded robust tree and its signature as if a view change is required for every transaction. However, in practice, the encoded tree and signature only need to be transmitted during the initial network setup or upon a view change. Adjusting for this, the actual \protocol{} overhead reduces by approximately 30 KB/min, resulting in a bandwidth usage closer to 162 KB/min, striking an efficient balance between performance and resource usage.

Mercury, focusing on propagation latency through its clustering-based approach and early outburst strategy, incurs an overhead of 322 KB/min. Mercury adds overhead through its Virtual Coordinate System (VCS) for topology maintenance and clustering-based propagation strategy, which involves metadata exchange and dynamic relaying to optimize latency.
Narwhal exhibits the highest bandwidth overhead at 730 KB/min, primarily due to its reliance on an intensive broadcast structure that requires collecting batch approvals from two-thirds of the network.

Several directions exist to further optimize \protocol{}'s performance. First, \protocol{} could manipulate batches of transactions. Then, an $(k{+}1, f{+}1{+}k)$ erasure coding scheme could divide a message into $f{+}1{+}k$ chunks, each one being disseminated over one of $f{+}1{+}k$ disjoint paths. A node would then receive at least $k{+}1$ chunks and recover the original batch of transactions. Depending on the blockchain, if specific roles are attributed to a subset of the nodes, e.g., validator nodes, then \protocol{} could be further optimized to minimize the transaction dissemination latency for these nodes.

\subsection{Role Distribution}

Figure~\ref{fig:depth} illustrates the distribution of roles among 200 nodes across 10 overlays. Each bar shows how often a node was assigned a given rank (depth). Rank 1 hereby corresponds to being an entry point. Disseminators start by transmitting through $f{+}1$ such entry points. The lengths of bars along the y-axis indicates how often each node had a specific rank in the 10 overlays.
In the robust tree structure, nodes closer to the entry points bear higher dissemination loads, while nodes near the leaves handle fewer messages. \protocol balances this load by varying node roles across overlays.
Since each robust tree has $f{+}1$ entry points, $10 \times (f{+}1)$ nodes are marked as such (i.e., in red; depth = 1).
Intermediate nodes make up the majority of the network. They are responsible for relaying messages from the entry points towards the leaves. The figure shows that ranks are widely distributed among nodes, ensuring that no specific node is consistently favored on the one side and also that no node is burdened with heavy forwarding responsibilities. Thus role rotation helps maintain balanced dissemination loads across the network.
The dissemination process ends at the leaf nodes. They receive messages but do not forward them.
The figure demonstrates that \protocol effectively achieves a balanced role distribution, with nodes taking different responsibilities across overlays, which ensures dissemination fairness and helps prevent front-running.

\subsection{Front-running Resistance.}
In the context of front-running resistance, \emph{dissemination fairness} is connected to how often malicious nodes are successful in manipulating the order with which transactions are received (front-running attacks). In this experiment, we evaluate the front-running resistance under different levels of malicious node activity. We simulate a network with varying percentages of malicious nodes (from 10\% to 33\%) that attempt to front-run transactions by observing and manipulating the mempool. A lower success rate indicates better fairness and resistance to manipulation.
The first malicious node to observe a victim transaction attempts to front-run it by generating and disseminating an adversarial transaction immediately in the network.
An attack succeeds if the adversarial transaction appears before the victim transaction in the blockchain. Note that under this definition we consider a front-running attack successful even if the adversarial transaction does not appear right before the victim transaction.

Figure~\ref{fig:front-run} shows that \protocol{} achieves the strongest protection against front-running attacks, with success rates as low as 2\% at 10\% malicious nodes and only 5.9\% at 33\%. This exceptional performance stems from \protocol{}' randomized dissemination mechanism, where the Threshold Random Seed (TRS) ensures unbiased overlay selection, making it exceedingly difficult for malicious nodes to predict and manipulate transaction propagation paths. Furthermore, \protocol{}' Robust Tree Encoding adds an additional layer of security by encoding dissemination structures in a verifiable manner, significantly mitigating manipulation opportunities.
\lz{} provides notable front-running resistance as well, with success rates ranging from 5\% at 10\% malicious nodes to 19\% at 33\%. Its mempool reconciliation mechanism ensures consistency across nodes, limiting adversaries' ability to exploit inconsistencies in transaction visibility. However, due to fewer mechanisms for randomized propagation compared to \protocol{}, \lz{} becomes progressively more vulnerable as the number of malicious nodes increases.
Narwhal demonstrates moderate resistance to front-running, with success rates increasing from 10\% at 10\% malicious nodes to 51\% at 33\%. It ensures the eventual propagation of transactions but lacks advanced defenses. This makes Narwhal more susceptible to manipulation as the percentage of malicious nodes increases, particularly under high adversarial activity.
Mercury exhibits the weakest front-running resistance among the protocols, with success rates escalating from 25\% at 10\% malicious nodes to 70\% at 33\%. While its clustering-based dissemination strategy offers efficiency, it makes the protocol vulnerable to manipulation by malicious nodes that can target critical nodes in the clusters to observe and reorder transactions. This centralized reliance on cluster leaders amplifies its susceptibility to front-running attacks under high adversarial conditions.

\subsection{Robustness.}

Robustness measures the likelihood that a message is successfully propagated to all honest nodes, even with varying percentages of Byzantine nodes disrupting propagation in a network of 10,000 nodes. Figure~\ref{fig:robustness} shows that \protocol{} achieves the highest robustness across all scenarios, with almost 99.9\% success rate at 10\% malicious nodes and maintaining 95\% even at 33\%.
This exceptional performance is attributed to its K-vertex-connected and robust overlay, which provides multiple redundant paths for message dissemination.
\lz{} follows with 97.5\% robustness at 10\% malicious nodes, though its performance drops to 80\% as adversarial influence increases. Its mempool reconciliation mechanism ensures consistency and decent robustness; however, its simpler dissemination structure with fewer redundant paths makes it less resilient under higher Byzantine node ratios.
Narwhal demonstrates similar robustness to \lz, achieving 95\% at 10\% malicious nodes and decreasing to 79\% at 33\%. Its reliable broadcast structure facilitates eventual message propagation but can experience delays or disruptions in the presence of adversarial behavior, reducing its overall consistency.
Mercury has the lowest robustness, starting at 89\% at 10\% malicious nodes and declining sharply to 55\% at 33\%. While its clustering-based approach optimizes for low latency, it introduces vulnerabilities where malicious clusters or failures of key nodes can lead to significant disruptions or network partitioning, highlighting the trade-off between low latency and robustness in its design.


\section{Conclusion} 
\label{sec:conclusions}

This paper presents \protocol{}, a fair, accountable, and low-latency dissemination protocol for use by the mempool subsystems of modern blockchain systems. Investigating front-running attacks that remain in contemporary mempools, we argued why fairness must be embedded in the lowest dissemination layer. \protocol{} achieves fairness by randomly selecting offline-optimized robust overlay structures to accelerate message dissemination, while forcing senders to adhere to the selected overlay and dissemination structure. At the costs of a slight bandwidth overhead of 192 KB/min (when compared to the 50 KB/min of the lightest baseline protocol), \protocol{} successfully mitigates front-running attacks, e.g., with less than 5.9\% of success rate in the presence of 33\% malicious nodes, at an achieved average communication latency of 83.22 ms (48\% of the latency of that protocol).
Directions for future work include investigating network hierarchies (e.g., for clustered consensus~\cite{yahyaoui2024tolerating,gupta2020resilientdb,thai2019hierarchical, amir2008steward}) and local overlay transformations that preserve fault-density despite churn.

\section*{Acknowledgments}
We would like to thank the anonymous reviewers for their valuable comments and advise.



%


\clearpage

\printbibliography{}

\end{document}